\documentclass[showpacs,prd,superscriptaddress,onecolumn]{revtex4}

\usepackage{amsmath}
\usepackage{amssymb}
\usepackage{amsfonts}
\usepackage{graphicx,bm}
\usepackage{dcolumn}
\usepackage[colorlinks=true]{hyperref}
\usepackage{color,amsxtra}
\usepackage{epsf}
\usepackage{enumerate}
\usepackage{hhline}
\usepackage{array}
\usepackage{tabularx}
\newcommand{\be}{\begin{equation}}
\newcommand{\ee}{\end{equation}}
\newcommand{\bea}{\begin{eqnarray}}
\newcommand{\eea}{\end{eqnarray}}
\newcommand{\beaa}{\begin{eqnarray*}}
\newcommand{\eeaa}{\end{eqnarray*}}

\newcommand{\nn}{\nonumber \\}

\begin{document}

\tolerance=5000

\title{Einstein-Gauss-Bonnet Cosmology Confronted with Observations}
\author{Sergei D. Odintsov}
\email{odintsov@ice.csic.es} \affiliation{Institut de Ci\`{e}ncies de l'Espai,
ICE/CSIC-IEEC, Campus UAB, Carrer de Can Magrans s/n, 08193 Bellaterra (Barcelona),
Spain}
 \affiliation{Instituci\'o Catalana de Recerca i Estudis Avan\c{c}ats (ICREA),
Passeig Luis Companys, 23, 08010 Barcelona, Spain}
 \author{V.K. Oikonomou}
\email{v.k.oikonomou1979@gmail.com;voikonomou@gapps.auth.gr}
\affiliation{Physics Department, Observatory, Aristotle University
of Thessaloniki}
\author{German~S.~Sharov}
 \email{sharov.gs@tversu.ru}
 \affiliation{Tver state university, Sadovyj per. 35, 170002 Tver, Russia}
 \affiliation{International Laboratory for Theoretical Cosmology,
Tomsk State University of Control Systems and Radioelectronics (TUSUR), 634050 Tomsk,
Russia}

\begin{abstract}
Several models within the framework of Einstein-Gauss-Bonnet
gravities are considered with regard their late-time
phenomenological viability. The models contain a non-minimally
coupled scalar field and satisfy a constraint on the scalar field
Gauss-Bonnet coupling, that guarantees that the speed of the
tensor perturbations is equal to the speed of light. The late-time
cosmological evolution of these Einstein-Gauss-Bonnet models is
confronted with the observational data including the Pantheon plus
Type Ia supernovae catalogue, the Hubble parameter measurements
(cosmic chronometers), data from cosmic microwave background
radiation (CMB) and  baryon  acoustic oscillations (BAO) including
the latest measurements from Dark Energy Spectroscopic Instrument
(DESI). Among the considered class of models some of them do not
fit the CMB and BAO data. However, there exists some models that
generate a viable Einstein-Gauss-Bonnet scenario with well-behaved
late-time cosmological evolution that fits the observational data
essentially better in comparison to the standard
$\Lambda$-Cold-Dark-Matter model.
\end{abstract}

\maketitle

\section{Introduction}

Theoretical cosmology has an elevated role in modern theoretical
physics. The observational data in the last two decades have been
cataclysmic offering modern cosmologists the possibility to find
suitable models for the description of, mainly, the late Universe,
and of course the early Universe. The observations originate from
the cosmic microwave background (CMB) radiation data
\cite{Planck2018}, but also from other sources, like the latest
Pantheon plus data \cite{PantheonP:2022} from Type Ia supernovae
(SNe Ia) and also from measurements from Dark Energy Spectroscopic
Instrument (DESI) \cite{DESI:2024} and finally data from baryon
acoustic oscillations (BAO). These data strongly challenge the
benchmark model for the description of the post-recombination
Universe, the $\Lambda$-Cold-Dark-Matter model ($\Lambda$CDM). On
the other hand, the $\Lambda$CDM model fits extremely well with
the CMB, but there seem to be strong tensions between the local
and global measurements of the Hubble constant, challenging the
$\Lambda$CDM. Indeed, there is a discrepancy of more than
$4\sigma$ between $\Lambda$CDM-based Planck estimations
\cite{Planck2018} of the Hubble constant and its measurements by
SH0ES group \cite{Riess2021}. Thus the $\Lambda$CDM is challenged
to say the least.

However, due to the incredible fitting of the $\Lambda$CDM with
the CMB, it seems that it is not a model to resent, but should be
a model to start with, keeping the good features of it, and
refining or eliminating the undesirable features of it. More
importantly, the shortcomings of the $\Lambda$CDM strongly
indicate one thing, that General Relativity (GR) is not able to
perfectly describe the post-recombination Universe. The
$\Lambda$CDM model is based on the foundations of GR, so the
tensions mean that GR needs to be modified. There are also other
issues that add up to this line of reasoning, for example the fact
that the Planck 2018 data \cite{Planck2018} allow the dark energy
equation of state to cross the phantom divide line, which simply
indicates that in terms of a GR theory one needs phantom scalar
fields to consistently produce a phantom equation of state. Also
the DESI 2024 data \cite{DESI:2024} indicate a dynamical dark
energy era, contrary to the $\Lambda$CDM evolution, a result that
is confirmed up to $4.2\sigma$ statistical confidence in the
remarkable findings of the 2025 DESI release \cite{DESI:2025zgx}.
Thus apparently GR is challenged and it is probable hereafter, not
possible, that extensions are needed to supplement it, for the
moment at large scales physics. There are however many theoretical
frameworks that emulate the $\Lambda$CDM model regarding the
late-time evolution, while being parallelized with the
observational data that challenge the $\Lambda$CDM model. Modified
gravity is the main proponent of this class of
$\Lambda$CDM-emulator theories, see
\cite{reviews1,reviews2,reviews3,reviews4} for reviews on the
subject. In the recent literature, it has been shown that one of
the most mainstream theories of modified gravity, namely $f(R)$
gravity, can fit the observational data much more successfully
compared to the $\Lambda$CDM model, see for example
\cite{OdintsovSGS:2024}, see also Refs.
\cite{Dai:2020rfo,He:2020zns,Nakai:2020oit,DiValentino:2020naf,Agrawal:2019dlm,Yang:2018euj,Ye:2020btb,Vagnozzi:2021tjv,
Desmond:2019ygn,OColgain:2018czj,Vagnozzi:2019ezj,
Krishnan:2020obg,Colgain:2019joh,Vagnozzi:2021gjh,Lee:2022cyh,Krishnan:2021dyb,Ye:2021iwa,Ye:2022afu,Verde:2019ivm,Menci:2024rbq,Adil:2023ara,
Reeves:2022aoi,Ferlito:2022mok,Vagnozzi:2021quy,DiValentino:2020evt,DiValentino:2019ffd,Sabogal:2024yha,Giare:2024smz}
for alternative viewpoints on this line of research. There are
several theories of modified gravity which are used and studied in
the literature, but only two stand out from a theoretical point of
view. These are $f(R)$ gravity theories, which contain Ricci
scalar corrections
\cite{Nojiri:2003ft,Capozziello:2005ku,Hwang:2001pu,Song:2006ej,Faulkner:2006ub,Olmo:2006eh,Sawicki:2007tf,Faraoni:2007yn,Carloni:2007yv,
Nojiri:2007as,Deruelle:2007pt,Appleby:2008tv,Dunsby:2010wg} and
Einstein-Gauss-Bonnet (EGB) theories containing the Gauss-Bonnet
(GB) invariant
\cite{Nojiri:2005jg,Cruz-DombrizS:2012,BenettiSCAL:2018,LeeT:2020,NavoE:2020,OdintsovOFF:2020,DeLaurentis:2015fea,Oikonomou:2017,
ElizaldeMOOF:2007,PavluchenkoT:2009,ChirkovGT:2021,Kawai:1998ab,AlexeyevTU:2000,Nojiri:2005vv,Nojiri:2006je,Cognola:2006,Bamba:2014zoa,
Guo:2009uk,Guo:2010jr,Jiang:2013gza,KohLLT:2014,Yi:2018gse,Odintsov:2018zhw,NojiriOOCP:2019,Odintsov:2019clh,Odintsov:2020sqy,
Odintsov:2020zkl,Odintsov:2020xji,Oikonomou:2020oil,Odintsov:2020mkz,NojiriOO:2023,OikonomouTR:2024,Oikonomou:2024etl,Fomin:2020,Pozdeeva:2020,
PozdeevaGSTV:2020,PozdeevaSTV:2024,OikonomouF:2020nm,PozdeevaSTV:2019,VernovP:2021,OdintsovOF:2021,OdintsovSGS:2022EGB,NojiriOP:2022}.
Both have a string theory or a quantum gravity origin. Indeed, it
is known that a scalar field in its vacuum configuration can
receive quantum corrections at one loop which contain Gauss-Bonnet
terms \cite{Codello:2015mba,Wheeler:1986} and $f(R)$ gravity
terms. In addition, both these theories can offer a unique
framework for which the unified description of the dark energy and
the inflationary era is realizable. The $f(R)$ gravity
confrontation with the observational data was presented in our
previous work \cite{OdintsovSGS:2024}, and in this work the focus
shall be on EGB theories. The GB term is a topological invariant
in four dimensions, so any linear term in the action turns out a
total derivative and thus the GB term does not affect the field
equations. But researchers suggested several ways to include the
GB term into a gravitational theory: (a) models with Lagrangians
nonlinear with respect to the Ricci scalar (modified GB gravities)
\cite{Nojiri:2005jg,Cruz-DombrizS:2012,BenettiSCAL:2018,LeeT:2020,NavoE:2020,OdintsovOFF:2020,DeLaurentis:2015fea,Oikonomou:2017};
(b) models with additional spatial dimensions
\cite{ElizaldeMOOF:2007,PavluchenkoT:2009,ChirkovGT:2021} and  (c)
theories including scalar fields coupled to the GB term
 \cite{Kawai:1998ab,AlexeyevTU:2000,Nojiri:2005vv,Nojiri:2006je,Cognola:2006,Bamba:2014zoa,
Guo:2009uk,Guo:2010jr,Jiang:2013gza,KohLLT:2014,Yi:2018gse,Odintsov:2018zhw,NojiriOOCP:2019,Odintsov:2019clh,Odintsov:2020sqy,
Odintsov:2020zkl,Odintsov:2020xji,Oikonomou:2020oil,Odintsov:2020mkz,NojiriOO:2023,OikonomouTR:2024,Oikonomou:2024etl,Fomin:2020,
Pozdeeva:2020,PozdeevaGSTV:2020,PozdeevaSTV:2024,OikonomouF:2020nm,PozdeevaSTV:2019,VernovP:2021,OdintsovOF:2021,OdintsovSGS:2022EGB,NojiriOP:2022}.
In this paper, we concentrate on this very last class of GB
theories with the GB term being coupled to a scalar field. These
models are actually called EGB theories or scalar GB theories
\cite{Nojiri:2005vv,Nojiri:2006je,Cognola:2006,Bamba:2014zoa,Guo:2009uk,Guo:2010jr,Jiang:2013gza,KohLLT:2014,
Yi:2018gse,Odintsov:2018zhw,NojiriOOCP:2019,Odintsov:2019clh,Odintsov:2020sqy,
Odintsov:2020zkl,Odintsov:2020xji,Oikonomou:2020oil,Odintsov:2020mkz,NojiriOO:2023,OikonomouTR:2024,Oikonomou:2024etl,Fomin:2020,
Pozdeeva:2020,PozdeevaGSTV:2020,PozdeevaSTV:2024,OikonomouF:2020nm,PozdeevaSTV:2019,VernovP:2021,OdintsovOF:2021,OdintsovSGS:2022EGB,NojiriOP:2022}.
These scenarios can potentially describe correctly both the early
and late-time era. Usually EGB theories are studied with regard to
their inflationary phenomenology
\cite{Guo:2009uk,Guo:2010jr,Jiang:2013gza,KohLLT:2014,Yi:2018gse,Odintsov:2018zhw,NojiriOOCP:2019,Odintsov:2019clh,Odintsov:2020sqy,
Odintsov:2020zkl,Odintsov:2020xji,Oikonomou:2020oil,Odintsov:2020mkz,Fomin:2020,Pozdeeva:2020,PozdeevaGSTV:2020,PozdeevaSTV:2024,OikonomouF:2020nm,
DeLaurentis:2015fea,Oikonomou:2017}. However, back in 2017 the
GW170817 event casted doubt on the theoretical consistency of EGB
theories, since these predict a speed of tensor perturbations
distinct from that of light. A refinement was proposed in the
literature by some of us, and specifically in the articles
\cite{Odintsov:2019clh,Odintsov:2020sqy,Odintsov:2020zkl,Odintsov:2020xji,Oikonomou:2020oil,Odintsov:2020mkz,NojiriOO:2023,OikonomouF:2020nm}
we considered  EGB gravities under an important constraint on the
coupling factor $\xi(\phi)$ connecting the scalar field $\phi$ and
the GB term: this constraint results from the condition of
equality the speed of gravitational waves and the speed of light.
This condition appears from the analysis of events with detected
gravitational waves (GW), in particular, the GW170817 neutron star
merger event by the LIGO/VIRGO collaboration
\cite{LIGOScientific:2017vwq} indicated that the GW speed should
nearly coincide with that of the light
\cite{Odintsov:2019clh,Odintsov:2020sqy}. In this article, we
consider a class of EGB model which belong to the class of
theories that align with the GW170817 event, studied and developed
in Refs.
\cite{Odintsov:2019clh,Odintsov:2020sqy,Odintsov:2020zkl,Odintsov:2020xji,Oikonomou:2020oil,Odintsov:2020mkz,NojiriOO:2023,OikonomouF:2020nm}.
We analyze some EGB models proposed in
Refs.~\cite{Oikonomou:2020oil,Odintsov:2020mkz,NojiriOO:2023,OikonomouF:2020nm,OdintsovOF:2021,OdintsovSGS:2022EGB}
and then concentrate on the most successful scenario and by using
observational datasets that include Pantheon+ SNe Ia catalogue,
CMB data, measurements of the Hubble parameter $H(z)$ (Cosmic
Chronometers) and DESI BAO data. The viable EGB model appears to
be essentially more successful than the standard $\Lambda$CDM
model in fitting this observational data.

The paper is organized as follows: the basic features and
equations of EGB gravity and the class of viable models are
considered in section \ref{Models}. In Sect.~\ref{Data}, the
observational data for SNe Ia, BAO, CMB and $H(z)$ is shown up as
well as the corresponding $\chi^2$ functions for fitting the
model. In Sect.~\ref{Viable} we select the most successful EGB
scenario and in the nest section describe its predictions and
compare them with the results of $\Lambda$CDM model. The
conclusions and final discussion are made in Sect.~\ref{Con}.

%%%%%%%%%%%%%%%%%%%%%%%%%%%%%%%%%%%%%%%%%%%%%%%%%%%%%%%%%%%%%%%%%%%%%%%%%%%%%%%%%%%
%wednesdayevening

\section{Essential Features of Einstein-Gauss-Bonnet Gravity}
\label{Models}

In this section we shall present the theoretical framework of
GW170817-compatible EGB gravity, the action of which is
\cite{OikonomouF:2020nm,PozdeevaSTV:2019,VernovP:2021,OdintsovOF:2021,OdintsovSGS:2022EGB},
\begin{equation}
 \label{action}
S=\int{d^4x\sqrt{-g}\left(\frac{R}{2\kappa^2}f(\phi)-\frac{1}{2}\partial_\mu\phi\partial^\mu\phi-V(\phi)-\xi(\phi)\,
 \mathcal{G}+\mathcal{L}_{(m)}\right)}\ ,
\end{equation}
where $\kappa^2=8\pi G$, $\phi$ is a scalar field, $R=R_\mu^\mu$
is the Ricci scalar, $\mathcal{L}_{(m)}$ is the matter Lagrangian
and $\mathcal{G}$ is the Gauss-Bonnet topological invariant
expressed via the Ricci tensor $R_{\mu\nu}$ and the Riemann
curvature tensor $R_{\mu\nu\sigma\rho}$. In many papers
\cite{Guo:2009uk,Guo:2010jr,Jiang:2013gza,KohLLT:2014,Yi:2018gse,Odintsov:2018zhw,NojiriOOCP:2019,Odintsov:2019clh,Odintsov:2020sqy,
Odintsov:2020zkl,Odintsov:2020xji,Oikonomou:2020oil,Odintsov:2020mkz,NojiriOO:2023,OikonomouTR:2024,Oikonomou:2024etl,Fomin:2020,Pozdeeva:2020,PozdeevaGSTV:2020,PozdeevaSTV:2024}
a more simple form of EBG gravity with the factor $f(\phi)=1$ is
explored,
\begin{equation}
 \label{action2}
S=\int{d^4x\sqrt{-g}\left(\frac{R}{2\kappa^2}-\frac{1}{2}\partial_\mu\phi\partial^\mu\phi-V(\phi)-\xi(\phi)\,
 \mathcal{G}+\mathcal{L}_{(m)}\right)}\ ,
\end{equation}
In this paper, we will work with both actions (\ref{action}) and
(\ref{action2}) in the framework of a flat
Friedman-Lema\^itre-Robertson-Walker (FLRW) metric
 \begin{equation}
\label{metric}
ds^2=-dt^2+a(t)^2\delta_{ij}dx^idx^j\ ,
 \end{equation}
where $a(t)$ is the scale factor, and for the FLRW spacetime, the
Ricci scalar and the Gauss-Bonnet invariant take the following
form,
  $$ %\be
   R=6(2H^2+\dot H),\qquad \mathcal{G}=24H^2(H^2+\dot H),\qquad H=\frac {\dot{a}}{a}\;.
  $$ %\ee
Varying the action (\ref{action}) with respect to the metric and
to $\phi$ we obtain the field equations
\cite{OikonomouF:2020nm,PozdeevaSTV:2019,VernovP:2021,OdintsovOF:2021,OdintsovSGS:2022EGB}:
\begin{eqnarray}
3f H^2&=&\kappa^2\bigg(\rho+\frac{1}{2}\dot\phi^2+V+24\dot\xi H^3\bigg)-3H\dot f\, ,
\label{eq1} \\
-2f \dot H&=&\kappa^2\bigg[\rho+P+\dot\phi^2-16\dot\xi H\dot H
-8H^2(\ddot\xi-H\dot\xi)\bigg] +\ddot f-H\dot f\, ,
\label{eq2}\\
\ddot\phi&+&3H\dot\phi+V'(\phi)-\frac{f^\prime(\phi)}{2\kappa^2}R+\xi'(\phi)\,\mathcal{G}=0\,
. \label{eqphi}
\end{eqnarray}
Here $\rho$ and $P$ are the matter density and pressure
respectively of the matter fluids, matter is the essential
component during the late-time epoch that is under consideration
in this research. Note that Eq. (\ref{eqphi}) is obtained by
combining the first two equations (\ref{eq1}) and (\ref{eq2}) and
may be omitted, but we will use it in our considerations.

In this paper, we impose the mentioned constraint on the coupling
function $\xi(\phi)$ coming from the equality of the GW speed and
speed of light, in order for the EGB theory to be
GW170817-compatible. This constraint is equivalent to the
following differential equation
 \cite{Odintsov:2019clh,Odintsov:2020sqy,Odintsov:2020zkl,Odintsov:2020xji,Oikonomou:2020oil,Odintsov:2020mkz,NojiriOO:2023,OikonomouF:2020nm}:
 \begin{equation}
 \ddot\xi=H\dot\xi\, ,
 \label{ddxi}  \end{equation}
 which can be easily integrated to be expressed in terms of the scale factor
 \begin{equation}
  \dot\xi=C\cdot a,\qquad\ C=\mbox{const}\;.
 \label{dxi}\end{equation}
For any concrete EGB scenario we should choose the factor
$f(\phi)$, the potential $V(\phi)$ and describe the matter content
$\rho$ and $P$. Then we can integrate the system of equations
(\ref{eq1})\,--\,(\ref{eqphi}) under the restrictions
(\ref{ddxi}), (\ref{dxi}) with given initial conditions at the
present time $t_0$. In particular, some EGB scenarios under the
restriction (\ref{ddxi}) were tested with the statefinder
parameter analysis in comparison to $\Lambda$CDM model for
late-time cosmology \cite{OdintsovOF:2021}. In
Ref.~\cite{OdintsovSGS:2022EGB} we explored some EGB scenarios
with Eq.~(\ref{ddxi}) solving the system
(\ref{eq1})\,--\,(\ref{eqphi}) in confrontation with late-time
observations. In that paper the matter components were assumed as
$\rho=\rho_m+\rho_r$, where $\rho_m$ describes baryons and cold
dark matter ($\rho_m\simeq\rho_b+\rho_c$) and $\rho_r$ refers to
relativistic particles (photons and neutrinos). For each matter
component its continuity equation,
 $$ %\begin{equation}
 \dot\rho_i+3H(\rho_i+P_i)=0
 $$ %\end{equation}
can be solved and we obtain,
\begin{equation}
\rho=\rho_{m}^0\left(a^{-3}+X_ra^{-4}\right)\, ,
 \label{rho}\end{equation}
where $X_r=\rho_{r}^0/\rho_{m}^0$, the index $0$ denotes present
day values (at $t=t_0$) and $a(t_0)=1$. To minimize the number of
free model parameters,  we fix the radiation to matter ratio as
estimated by Planck \cite{Planck:2013,Odintsov:2020voa}:
\begin{equation}
X_r=\frac{\rho_r^0}{\rho_m^0}=2.9656\times 10^{-4}\,. \label{Xrm}
\end{equation}
Models from the considered class can be viable if (a) they allow physical solutions and
(b) these solutions satisfy observational limitations. In the next section we describe
actual observational data used in this paper.

%%%%%%%%%%%%%%%%%%%%%%%%%%%%%%%%%%%%
\section{Observational data} \label{Data}
%%%%%%%%%%%%%%%%%%%%%%%%%%%%%%%%%%%%

We test the EGB models (\ref{action}), (\ref{action2}) with observational data including
the Pantheon+ sample database \cite{PantheonP:2022} of Type Ia Supernovae (SNe Ia), the
Hubble parameter $H(z)$ measurements, observable parameters of Cosmic Microwave
Background radiation (CMB) and Baryon Acoustic Oscillations (BAO) with the latest BAO
measurements from Dark Energy Spectroscopic Instrument (DESI) collaboration
\cite{DESI:2024}. To confront any model with these observations we use some methods
developed previously in papers
\cite{Odintsov:2020voa,Odintsov:2017qif,OdintsovSGS_LnAx:2024}).\\

For SNe Ia we use $N_{\mathrm{SN}}=1701$ Pantheon+  datapoints \cite{PantheonP:2022}
with information of the distance moduli $\mu_i^\mathrm{obs}$ at redshifts $z_i$ from
1550 SNe Ia and compute the $\chi^2$ function:
 \bea
\chi^2_{\mathrm{SN}}(\theta_1,\dots)&=&\min\limits_{H_0} \sum_{i,j=1}^{N_\mathrm{SN}}
 \Delta\mu_i\big(C_{\mathrm{SN}}^{-1}\big)_{ij} \Delta\mu_j\ ,\nn
 \Delta\mu_i&=&\mu^\mathrm{th}(z_i,\theta_1,\dots)-\mu^\mathrm{obs}_i\ .
 \label{chiSN}\eea
Here $\theta_j$ are free model parameters, the Hubble constant
$H_0$ is considered  as a nuisance parameter,
$C_{\mbox{\scriptsize SN}}$ is the $N_{\mathrm{SN}}\times
N_{\mathrm{SN}}$ covariance matrix and $\mu^\mathrm{th}$ are the
theoretical values for the distance moduli depending on redshift
$z=a^{-1}-1$:
\begin{equation}
 \mu^\mathrm{th}(z) = 5 \log_{10} \frac{(1+z)\,D_M(z)}{10\mbox{pc}},\qquad D_M(z)= c \int\limits_0^z\frac{d\tilde z}{H(\tilde
 z)}.    \label{muDM}
\end{equation}
In this paper, we use the Hubble parameter $H(z)$ measurements named ``Cosmic
Chronometers'' (CC): they are made by the method of different ages $\Delta t$ for
galaxies with known variations of redshifts $\Delta z$. The values $H(z)$ are estimated
via the relation:
$$ %\begin{eqnarray}
H (z)= \frac{\dot{a}}{a} \simeq -\frac{1}{1+z} \frac{\Delta z}{\Delta t}\,.
$$
For the present analysis $N_H=32$ data points of
$H^\mathrm{obs}(z_i)$ are used with references in the previous
papers \cite{OdintsovSGS_LnAx:2024}). The $\chi^2$ function for
$H(z)$ fittings yields:
\begin{equation}
\chi^2_{H}= \sum_{i=1}^{N_H} \left[\frac{H^\mathrm{obs}(z_i)
 -H^\mathrm{th}(z_i; \theta_k)}{\sigma_{H,i}}\right]^2 \, .
\label{chiH}
\end{equation}
CMB observational parameters are related to the photon-decoupling
epoch at redshifts close to $z_*=1089.80 \pm0.21$ and are given by
the vector \cite{Planck2018}:
\begin{equation}
\mathbf{x}=\left(R,\ell_A,\omega_b \right)\, ,\quad
R=\sqrt{\Omega_m^0}\frac{H_0D_M(z_*)}c\, ,\quad \ell_A=\frac{\pi D_M(z_*)}{r_s(z_*)}\, ,
\quad\omega_b=\Omega_b^0h^2
 \label{CMB} \end{equation}
whose  estimations are \cite{ChenHW2018}:
\begin{equation}
\mathbf{x}^\mathrm{Pl}=\left( R^\mathrm{Pl},\ell_A^\mathrm{Pl},\omega_b^\mathrm{Pl}
\right) =\left( 1.7428\pm0.0053,\;301.406\pm0.090,\;0.02259\pm0.00017 \right) \, .
\label{CMBpriors}
\end{equation}
The comoving sound horizon  $r_s(z_*)$ is obtained by the integral
\cite{OdintsovSGS_LnAx:2024}):
  \begin{equation}
r_s(z)=  \int_z^{\infty} \frac{c_s(\tilde z)}{H (\tilde z)}\,d\tilde
z=\frac1{\sqrt{3}}\int_0^{1/(1+z)}\frac{da}
 {a^2H(a)\sqrt{1+\big[3\Omega_b^0/(4\Omega_\gamma^0)\big]a}}\ ,
  \label{rs2}\end{equation}
with the estimation of $z_*$ given in Refs.~\cite{ChenHW2018}. The
reduced baryon fraction $\omega_b$ is considered as the nuisance
parameter in the following $\chi^2$ function:
\begin{equation}
\chi^2_{\mathrm{CMB}}=\min_{\omega_b,H_0}\Delta\mathbf{x}\cdot
C_{\mathrm{CMB}}^{-1}\left( \Delta\mathbf{x} \right)^{T}\, ,\quad \Delta
\mathbf{x}=\mathbf{x}-\mathbf{x}^\mathrm{Pl}\,, \label{chiCMB}
\end{equation}
where $C_{\mathrm{CMB}}=\| \tilde C_{ij}\sigma_i\sigma_j \|$ is
the covariance \cite{Odintsov:2020voa,ChenHW2018}.  For the Baryon
Acoustic Oscillations (BAO) we consider new data from Dark Energy
Spectroscopic Instrument (DESI) Data Release 1 \cite{DESI:2024} is
considered. We calculate and compare with measurements the value
\begin{equation}
d_z(z)= \frac{r_s(z_d)}{D_V(z)}\, ,\qquad D_V(z)=\bigg[\frac{cz D_M^2(z)}{H(z)}
\bigg]^{1/3}\, , \label{dzDV}
\end{equation}
where $z_d$ being the redshift at the end of the baryon drag era whereas the comoving
sound horizon $r_s(z)$ is obtained as the integral (\ref{rs2}). The estimations for
$z_d$ and for the ratio of baryons and photons $\Omega_b^0/\Omega_\gamma$ are fixed by
the Planck 2018 data \cite{Planck2018}.

In this paper, we use  6 BAO datapoints, shown in Table~\ref{DESI} and provided by DESI
DR1 data \cite{DESI:2024}. These measurements include BAO data from clustering of
galaxies, quasars and the Lyman-$\alpha$ forest in the redshift range $0.1<z<4.16$. The
$\chi^2$ function is
\begin{equation}
\chi^2_{\mathrm{BAO}}(\theta_1,\dots)=\sum_{i=1}^{6}
\left[\frac{d_z^\mathrm{obs}(z_i)-d_z^\mathrm{th}(z_i,\dots)}{\sigma_{d_z,i}}\right]^2
\, . \label{chiBAO}
\end{equation}
\begin{widetext}
\begin{table}[hb]
%\begin{center}
\begin{tabular}{|c|c|c|c|c|c|c|}
\hline  $z_\mathrm{eff}$ & 0.295& 0.51 & 0.706 & 0.93 & 1.317 & 2.33\\
\hline
 $z$ range &0.1 - 0.4&0.4 - 0.6&0.5 - 0.8&0.8 - 1.1&1.1 - 1.6 &1.77 - 4.16\\
\hline
 $d_z$ &$0.1261 \pm0.0024  $&$0.0796 \pm0.0018  $ &$0.0629 \pm0.0014  $ &$0.05034\pm0.0008  $
&$0.04144\pm0.0011  $ &$0.03173\pm 0.00073$\\
\hline
 \end{tabular}
 \caption{DESI DR1 BAO data.}
%\end{center}
\label{DESI}
\end{table}
\end{widetext}
Then, the free parameters for a given EGB model should be fitted
with all these sources of data. In the next section, we estimate
viability of some EGB scenarios in this test.
\section{Viable EGB Models}
\label{Viable}

We begin our analysis from the most popular EGB models with the action (\ref{action2})
\cite{Odintsov:2018zhw,NojiriOOCP:2019,Odintsov:2019clh,Odintsov:2020sqy,
Odintsov:2020zkl,Odintsov:2020xji,Oikonomou:2020oil,Odintsov:2020mkz,NojiriOO:2023,OikonomouTR:2024,Oikonomou:2024etl}.
Their dynamics will be determined if we impose the condition (\ref{ddxi}) or $\dot\xi=C
a$ and fix the potential $V(\phi)$. In the mentioned papers the authors explore
different variants of $V(\phi)$ having a preference for power-law and exponential
potentials that may be written in the following common form:
 \begin{eqnarray}
V(\phi)&=&V_0\bigg(\frac\phi{\phi_0}\bigg)^\alpha,\label{Vpower} \\
V(\phi)&=&V_0\exp\big[\tilde\alpha(\phi-\phi_0)\big]\,. \label{Vexp}
\end{eqnarray}
Recall that the index ``$0$'' denotes the present day values of
variables. Instead of the Hubble parameter $H(t)$, the scalar
$\phi(t)$, its derivative $\dot\phi$, the constants $\rho_m^0$,
$V_0$, $C$ it is convenient to use dimensionless variables and
parameters:
 \begin{equation}
 E=\frac H{H_0},\quad\; \varphi=\kappa\phi,\quad\; \psi=\frac{\kappa}{H_0}\dot\phi;
 \qquad\Omega_m^0=\frac{\kappa^2\rho_m^0}{3H_0^2}, \quad\;\Omega_V=\frac{\kappa^2V_0}{3H_0^2} ,
 \quad\;\lambda=\kappa^2H_0 C\,.
 \label{dimen}
 \end{equation}
 In this notation the potentials  (\ref{Vpower}) and (\ref{Vexp}) take the common form
 $$V(\varphi)=V_0\cdot v(\varphi),\qquad  v(\varphi)=(\varphi/\varphi_0)^\alpha\quad\mbox{or}
 \quad  v(\varphi)=\exp\big[\alpha(\varphi-\varphi_0)\big]
 $$
the condition (\ref{dxi}) transforms into
$\dot\xi=\frac\lambda{\kappa^2H_0}a$, and the system of equations
(\ref{eq1})\,--\,(\ref{eqphi}) for the considered case
(\ref{action2}) ($f=1$) can be expressed as:
 \begin{eqnarray}
E^2(1-8\lambda aE)&=&\Omega_m^0\big(a^{-3}+X_ra^{-4}\big)+
\tfrac{1}{6}\psi^2+\Omega_V v(\varphi)\, ,\label{eq12} \\
2E\frac{dE}{dx}\big(1-8\lambda
aE\big)&=&-\Omega_m^0\big(3a^{-3}+4X_ra^{-4}\big)-\psi^2\,,
\label{eq22}\\
\frac{d\psi}{dx}+3\psi&+&3\frac{\Omega_Vv'(\varphi)}E+24\frac{\lambda a}{\psi}
E^2\bigg(E+\frac{dE}{dx}\bigg)=0\, . \label{eqphi2}
\end{eqnarray}
Here $x=\log a$, $\frac d{dt}=H\frac d{dx}$, $\psi=E\frac{d\varphi}{dx}$. This system
can be solved numerically in the ``past'' direction starting  at the present time as
described in Ref.~\cite{OdintsovSGS:2022EGB}. We fix initial conditions at the present
time $t=t_0$ (equivalent to $x=0$ or $a=1$), where the following free model parameters
should be fixed:
 \be
E\big|_{x=0}=1,\qquad \varphi\big|_{x=0}=\varphi_0,\qquad
\psi\big|_{x=0}=\psi_0=\pm\sqrt{6\big[1-8\lambda-\Omega_m^0(1+X_r)-\Omega_V\big]}\ .
 \label{phsi0} \ee
For any set of free model parameters $\Omega^0_m$, $\Omega_V$, $H_0$, $\lambda$,
$\varphi_0$, $\alpha$ for both power-law (\ref{Vpower}) and exponential (\ref{Vexp})
potentials we can obtain the solution $E(x)$, $\varphi(x)$ of the system
(\ref{eq12})\,--\,(\ref{eqphi2}) and (if this solution is physical) compare it with
observational data. For this purpose we use the Hubble parameter $E(z)$ or
$H(z)=H_0E(z)$,  where $z=a^{-1}-1=e^{-x}-1$.

\begin{figure}[th]
\centerline{ \includegraphics[scale=0.66,trim=5mm 0mm 2mm
-1mm]{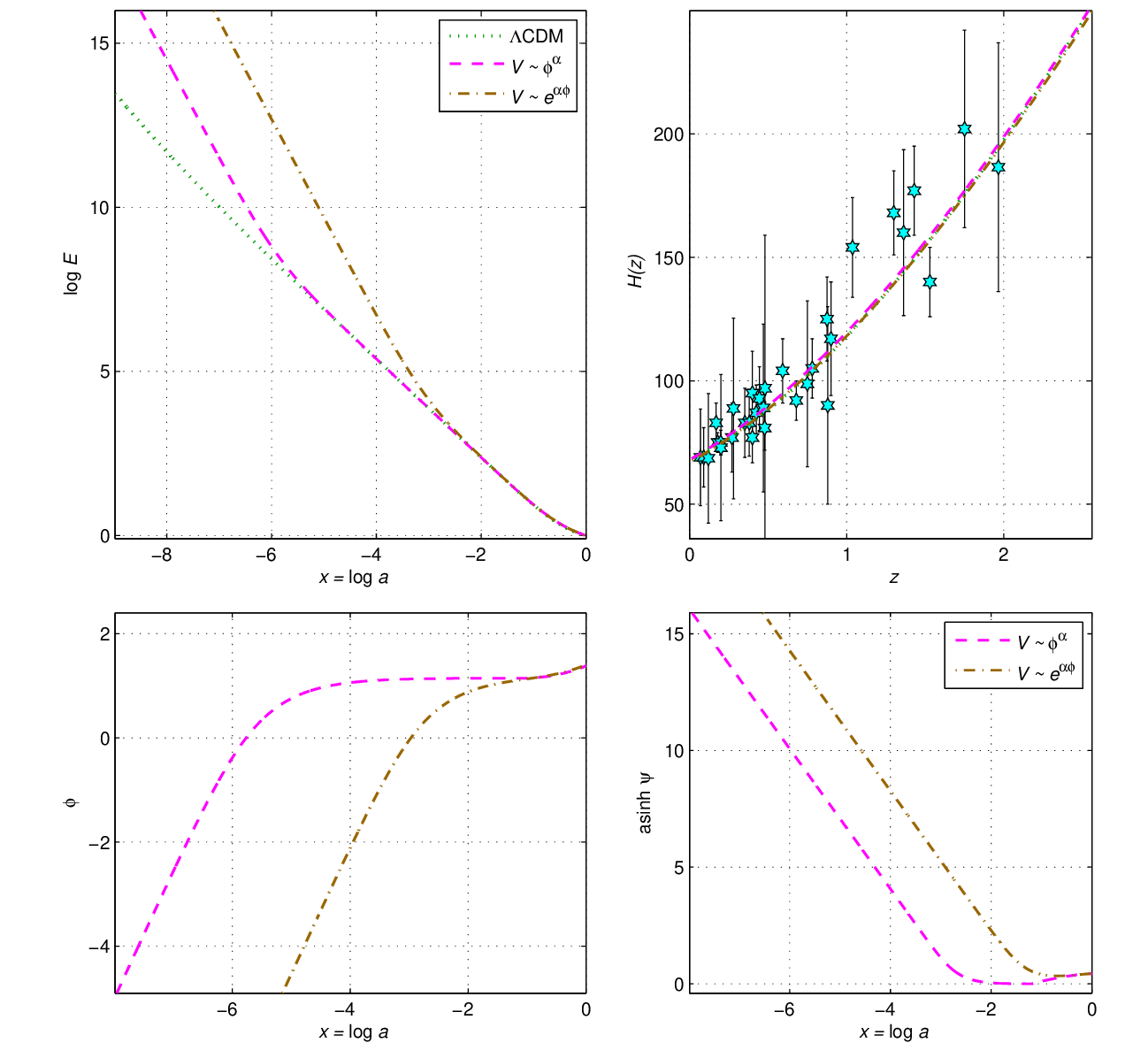}} \caption{Evolution of the normalized Hubble
parameter $E=H/{H_0}$ (top-left panel), the Hubble parameter
$H(z)$ at small redshifts (top-right panel), the scalar field
$\varphi=\kappa\phi$ and its derivative
$\psi=E\frac{d}{dx}\varphi$ (bottom panels) versus $x=\log a$ for
the EGB models with the power-law and exponential potentials
(\ref{Vpower}), (\ref{Vexp}) in comparison with the $\Lambda$CDM
model (\ref{LCDM}). } \label{F1}
\end{figure}

Our calculations showed that physical solutions describing
Pantheon+ SNe Ia data \cite{PantheonP:2022} and CC $H(z)$ data
exist for both power-law and exponential models (\ref{Vpower}),
(\ref{Vexp}). If we consider only  $\chi^2$ functions for  SNe Ia
(\ref{chiSN}) and CC (\ref{chiH}) or their sum
$\chi^2_{\mathrm{SN}}+\chi^2_H$, these EGB models appear to be
more successful than the standard $\Lambda$CDM model with,
 \begin{equation}
\frac{H(z)}{H_0}=\big[\Omega_m^0(a^{-3}+X_ra^{-4})+\Omega_\Lambda\big]^{1/2},\qquad
\Omega_\Lambda=1-\Omega_m^0(1+X_r)\;.
 \label{LCDM}  \end{equation}
The mentioned Pantheon+ SNe Ia and $H(z)$ observations are related
with the late-time epoch at the range $0<z<2.5$. However, at more
early times (at $z>100$) for both EGB scenarios (\ref{Vpower}) and
(\ref{Vexp}) the normalized Hubble parameter $E(x)$ deviates from
the $\Lambda$CDM expression (\ref{LCDM}). Such a behavior can be
seen in the top-left panel of Fig.~\ref{F1} for both  EGB
potentials (\ref{Vpower}) and (\ref{Vexp}). On the one hand the
$\Lambda$CDM normalized Hubble parameter (\ref{LCDM}) behaves as
$E\sim a^{-2}$ before the recombination epoch and
$E\simeq\sqrt{\Omega_m^0}\cdot a^{-3/2}$ after (at $10^3>z\gg1$ or
$-7>x>-3$); on the other hand both EGB scenarios at high redshifts
grow quickly in the power-law manner: $E\sim a^{-3}$. Such a
behavior is the consequence of growing value $\psi\sim a^{-3}$ at
high redshifts in accordance with Eq.~(\ref{eqphi2}) that can be
seen in the bottom-right panel  of Fig.~\ref{F1}. As the result
the term $\frac{1}{6}\psi^2$ becomes dominating in the right hand
side of equation $(\ref{eq12})$ that leads to the mentioned growth
$E\sim a^{-3}$.
The examples in the top-right panel of  Fig.~\ref{F1} show that
both EGB scenarios (\ref{Vpower}) and (\ref{Vexp}) can describe CC
$H(z)$ data (32 CC datapoints are depicted). This also concerns to
Pantheon+ SNe Ia data at the same redshift range $0<z<2.5$ and we
can achieve better fits for the sum
$\chi^2_{\mathrm{SN}}+\chi^2_H$ compared to the $\Lambda$CDM
model. In particular, in Fig.~\ref{F1} the model parameters are
chosen near the best fits: $ \Omega^0_m=0.286$, $H_0=68.0$
km\,(s\,Mpc)${}^{-1}$ for all models; $\Omega_V=0.68$,
$\varphi_0=1.4$ for both EGB scenarios; $\alpha=-1$,
$\lambda=-5\cdot10^{-6}$ for the power-law and $\alpha=-0.75$,
$\lambda=-3\cdot10^{-6}$ for the exponential model.

The above described asymptotic behavior $E\sim a^{-3}$ for both
EGB scenarios (\ref{Vpower}) and (\ref{Vexp}) takes place for any
set of free parameters and initial conditions. This said behavior
leads to an essential difference in the calculated integral
(\ref{rs2}) for the comoving sound horizon  $r_s(z)$ between these
EGB scenarios and the $\Lambda$CDM model. The value $r_s(z)$
determines not only theoretical CMB parameter $\ell_A$ (\ref{CMB})
related to the recombination epoch, but also the BAO parameter
$d_z$ (\ref{dzDV}). As the result for EGB scenarios (\ref{Vpower})
and (\ref{Vexp}) we have very large discrepancies between
theoretical and observational CMB and BAO parameters leading to
high values of $\chi^2_{\mathrm{BAO}}$ and
$\chi^2_{\mathrm{CMB}}$: minima of these $\chi^2$ functions and
their sum for the mentioned EGB models exceed the corresponding
$\Lambda$CDM minima in many orders of magnitude.

Due to this reason the EGB scenarios with the action (\ref{action2}), the condition
(\ref{dxi}) and potentials (\ref{Vpower}), (\ref{Vexp}) (and other physically acceptable
potentials) concede to $\Lambda$CDM model when we test them with the whole set of SNe
Ia, $H(z)$, CMB and BAO observational data. However, among a more wide class of EGB
models with the action (\ref{action}) we can find scenarios with suitable behavior of
the Hubble parameter $H(z)$ at high redshifts. In particular, we can consider the model
of the type (\ref{action}) with the factor $f(\phi)=\phi_0/\phi=\varphi_0/\varphi$
explored in the papers \cite{OdintsovOF:2021,OdintsovSGS:2022EGB}. For this model with
the corresponding power-law potential $V(\phi)\sim\phi^{-1}$ we use the following terms
in the Lagrangian (\ref{action}):
\begin{equation}
\frac{R}{2\kappa^2}f(\phi)-V(\phi)=\frac{R}{2\kappa^2}\frac{\varphi_0}\varphi
-V_0\frac{\varphi_0}\varphi\;.
 \label{Mod3}
\end{equation}
For the model (\ref{Mod3}) under the condition (\ref{ddxi}) or (\ref{dxi}) (restricting
the GW speed) in notation (\ref{dimen}) the equations (\ref{eq1})\,--\,(\ref{eqphi}) are
transformed into the following system:
\begin{eqnarray}
E^2\big[f(\phi)-8\lambda
aE\big]&=&\Omega_m^0\big(a^{-3}+X_ra^{-4}\big)+\tfrac{1}{6}\psi^2+(\Omega_V+E\psi/\varphi)\,f(\phi)\,,
\label{eq13} \\
2E\frac{dE}{dx}\bigg(\frac{\varphi_0}\varphi-8\lambda
aE\bigg)&=&-\Omega_m^0\big(3a^{-3}+4X_ra^{-4}\big)-\psi^2+
\frac{\varphi_0}{\varphi^2}\bigg(E\frac{d\psi}{dx}-E\psi-2\frac{\psi^2}{\varphi} \bigg),
\label{eq23}\\
\frac{d\psi}{dx}+3\psi&-&3\frac{\Omega_V\varphi_0}{E\varphi^2}+24\frac{\lambda a}{\psi}
E^2\bigg(E+\frac{dE}{dx}\bigg)+3\frac{\varphi_0}{\varphi^2}\bigg(2E+\frac{dE}{dx}\bigg)=0\,
. \label{eqphi3}
\end{eqnarray}
This system is solved numerically similarly to Eqs.~(\ref{eq12})\,--\,(\ref{eqphi2})
with the initial conditions similar to (\ref{phsi0}): $E\big|_{x=0}=H_0/H_0=1$,
$\varphi\big|_{x=0}=\varphi_0$,  but with $\psi_0$ determined by Eq.~(\ref{eq13}) at
$x=0$:
 \be
\psi\big|_{x=0}=
-3\varphi_0^{-1}+\sqrt{9\varphi_0^{-2}-6\big[\Omega_m^0(1+X_r)+\Omega_V-1+8\lambda\big]}\ .
 \ee
For deriving the solution at every step we calculate $\psi(x)$ and
the preliminary value $\varphi(x)=\int\psi E^{-1}dx$, then
determine $E$ by solving the equation (\ref{eq13}) and redefine
$\varphi(x)$. Further, we extract the  derivative
$\frac{d\psi}{dx}$ from equations (\ref{eq23}) and (\ref{eqphi3})
in order to calculate $\psi(x-\Delta x)$ at the nest step.

In these calculations we obtained solutions  for the EGB model
(\ref{Mod3}) behaving closely to $\Lambda$CDM solutions
(\ref{LCDM}) at late and early times. Our results may be
confronted with the observational data,  and we present this
analysis in the following section.

%%%%%%%%%%%%%%%%%%%%%%%%%%%%%%%%%%%%%%%%%%
\section{Further Confrontation of Viable EGB with the Observational Data}
\label{Results}
%%%%%%%%%%%%%%%%%%%%%%%%%%%%%%%%%%%%%

\begin{figure}[ht]
\centerline{ \includegraphics[scale=0.62,trim=5mm 0mm 2mm -1mm]{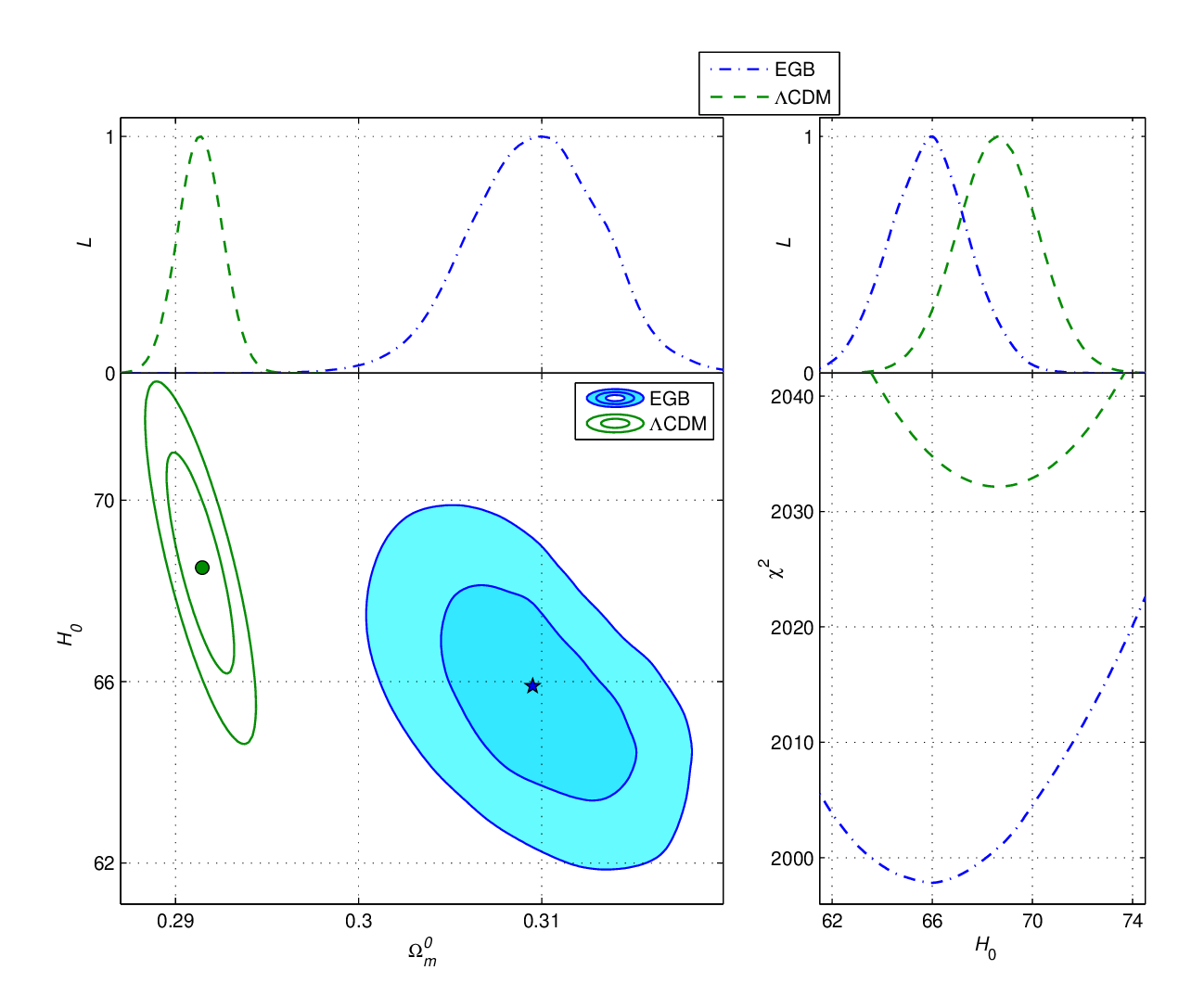}}
 \caption{Contours plots at $1\sigma$ and $2\sigma$ CL in the $\Omega_m^0-H_0$ plane,
likelihoods $\mathcal{L}(\Omega_m^0)$, $\mathcal{L}(H_0)$ B and one-parameter
distributions $\chi^2(H_0)$ for the EGB model (\ref{Mod3}) in comparison with
$\Lambda$CDM model. }
  \label{F2}
\end{figure}

\begin{figure}[ht]
\centerline{ \includegraphics[scale=0.62,trim=5mm 0mm 2mm -1mm]{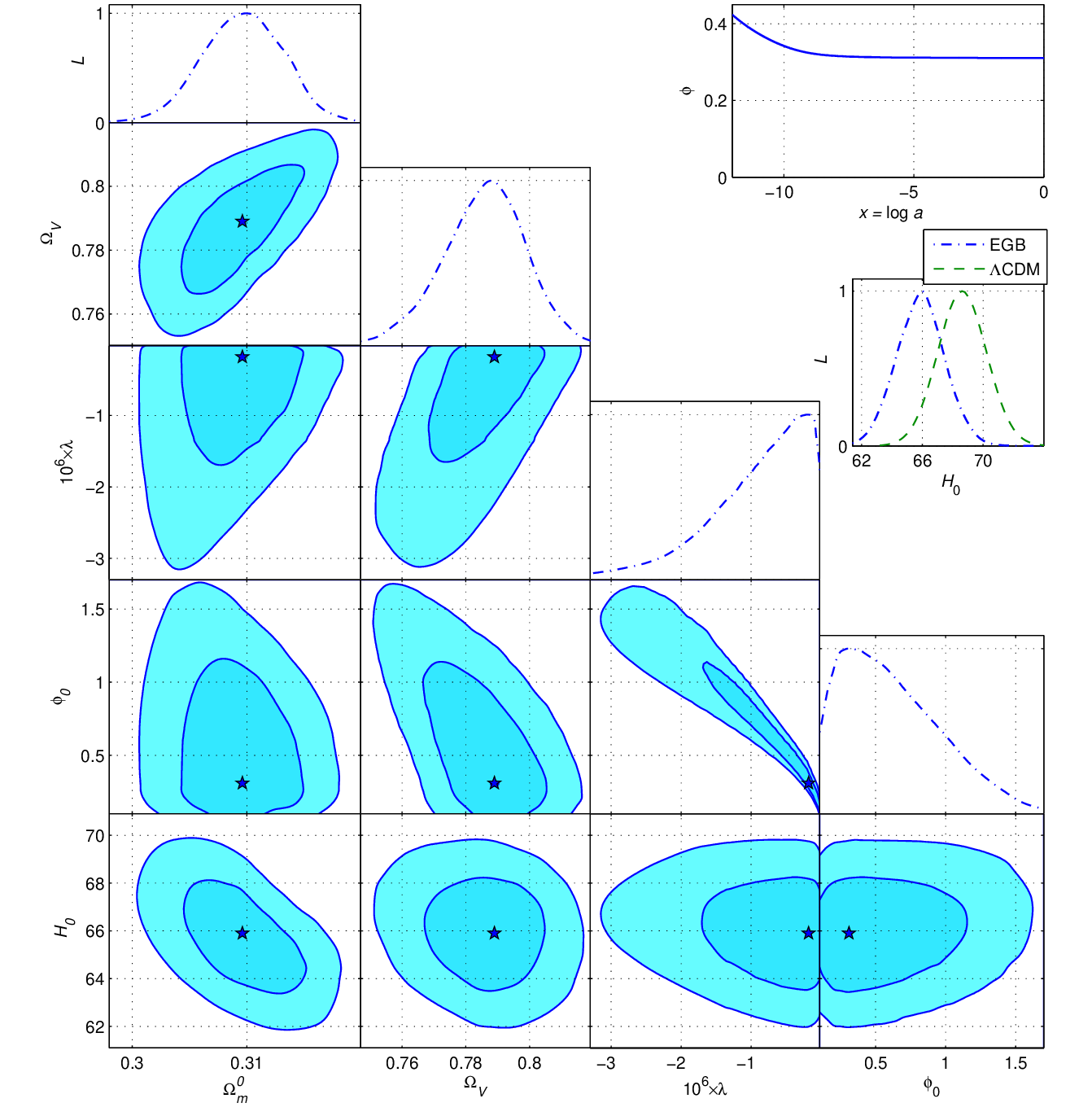}}
 \caption{EGB model (\ref{Mod3}): $1\sigma$ and $2\sigma$ contours plots, likelihoods for all
 free parameters and evolution of the scalar field $\varphi$ in the top-right panel. }
  \label{F3}
\end{figure}

To compare the EGB model (\ref{Mod3}) with the observational datasets described in
Sect.~\ref{Data} we fit its free parameters
 \begin{equation}
 \Omega^0_m,\quad \Omega_V,\quad H_0,\quad\lambda,\quad\varphi_0\,,
\label{5param}\end{equation} calculating the corresponding
$\chi^2_i$ functions (\ref{chiSN}), (\ref{chiH}), (\ref{chiCMB}),
(\ref{chiBAO}) for  SNe Ia, $H(z)$, CMB and BAO datasets and the
total (summarized) $\chi^2_{\mathrm{tot}}$ function:
\begin{equation}
\chi^2\equiv\chi^2_{\mathrm{tot}}=\chi^2_{\mathrm{SN}}+\chi^2_H+\chi^2_{\mathrm{BAO}}+\chi^2_{\mathrm{CMB}}\;.
 \label{chitot}
\end{equation}
We calculated this  $\chi^2$  function  for each pair of the free parameters
(\ref{5param}) and presented the results in Fig.~\ref{F2} and Fig.~\ref{F3} with the
contour plots at $1\sigma$ (68.27\%) and $2\sigma$ (95.45\%) confidence levels (CL) for
two-parameter distributions $\chi^2(\theta_j,\theta_k)$.

In Fig.~\ref{F2} we compare the EGB model (\ref{Mod3}) under the
condition (\ref{dxi}) with the $\Lambda$CDM model (\ref{LCDM}). In
the bottom-left panel $1\sigma$ and $2\sigma$ contours plots are
depicted in the plane $\Omega_m^0-H_0$ of their common parameters
for two-parameter distributions. In the EGB case this distribution
is the following minimum over other free parameters:
 $$ %\be
 \chi^2(\Omega_m^0,H_0)=\min\limits_{\Omega_V,\lambda,\varphi_0} \chi^2(\Omega_m^0,\Omega_V,\lambda,\varphi_0,H_0)\,.
 $$ %\ee
The minimum points for $\chi^2(\Omega_m^0,H_0)$ in this panel are
labelled with the star for EGB (filled contours) and with the
circle for $\Lambda$CDM model.
In the bottom-right panel of Fig.~\ref{F2} the one-parameter
distributions $\chi^2(H_0)$ (minimized over the rest of the free
parameters) for both models are depicted and compared; in the top
panels one can see the corresponding likelihoods
$\mathcal{L}(\Omega_m^0)$ and $\mathcal{L}(H_0)$, where
\begin{equation}
\mathcal{L}(\theta_k)\propto \exp\left[-\frac12\chi^2(\theta_k)\right]\,.
\label{likelihood}
\end{equation}
The contours and the predicted best fits for the EGB and $\Lambda$CDM models strongly
differ in the bottom-left panel of Fig.~\ref{F2}: this difference exceeds $3\sigma$.
These best fits deviate especially sharply for the parameter $\Omega_m^0$:
$\Omega_m^0=0.2913^{+0.0012}_{-0.0012}$ for $\Lambda$CDM and
$\Omega_m^0=0.3096^{+0.0040}_{-0.0036}$ for EGB.
 These and other best fits and $1\sigma$ errors for the free parameters
$\theta_k$ are tabulated in Table~\ref{Estim}, they are obtained from one-parameter
distributions $\chi^2(\theta_k)$.  From this table and likelihoods $\mathcal{L}(H_0)$ in
the top-right panel of Fig.~\ref{F2} one may conclude that the best fits for the Hubble
constant $H_0$ for both models differ of  order $1\sigma$ and the EGB model predicts the
lowest estimation: $H_0=65.90^{+1.48}_{-1.54}$ km\,/(s\,Mpc).

In the bottom-right panel of Fig.~\ref{F2} the one-parameter distributions $\chi^2(H_0)$
illustrate the large difference between two considered models: the EGB model
(\ref{Mod3}) appeared to be much more successful in its absolute minimum
$\min\chi^2\simeq1997.76$ in comparison with the $\Lambda$CDM result
$\min\chi^2\simeq2032.15$. One may conclude that the EGB model fits essentially better
the observational data in comparison to $\Lambda$CDM model according to the minimum of
$\chi^2$.

Moreover, in order to improve our statistical analysis we can take
into account the large number $N_p=5$ of free parameters
(\ref{5param}) for the EGB model, while $\Lambda$CDM just contains
$N_p=2$, and consider the Akaike information criterion (AIC
parameter) \cite{Odintsov:2020voa}:
 \begin{equation}
\mathrm{AIC} = \min\chi^2 +2N_p
 \label{AIC} \end{equation}
 and Bayesian information criterion (BIC) \cite{Liddle_ABIC:2007}
  \begin{equation}
\mathrm{BIC} = \min\chi^2 +N_p\cdot\log(N_d)\;,
 \label{BIC} \end{equation}
 where $N_d$ is the number of data points. In Table~\ref{Estim} one can see that the advantage of the EGB
model (\ref{Mod3}) in $\min\chi^2$ remains at a high level when we
consider both  Akaike and Bayesian information criteria. In
particular, we see the difference between these models
$\Delta\mbox{AIC}=\mbox{AIC}_\mathrm{EGB}-\mbox{AIC}_\mathrm{{\Lambda}CDM}=-28.39$
and
$\Delta\mbox{BIC}=\mbox{BIC}_\mathrm{EGB}-\mbox{BIC}_\mathrm{{\Lambda}CDM}=-12.002$
both in favor of the EGB model (\ref{Mod3}).
\begin{table}[bh]
 \centering
 {\begin{tabular}{|l|c|c|c|c|c|c|c|c||}  \hline
 \hline  Model &   $\min\chi^2/d.o.f$& AIC & BIC& $\Omega_m^0$& $H_0$& $\Omega_V$& $10^6\cdot\lambda$& $\varphi_0$ \\
\hline
 EGB (\ref{Mod3}) & 1997.76 /1738 & 2007.76&$2035.074$&$0.3096^{+0.0040}_{-0.0036}$
& $65.90^{+1.48}_{-1.54}$ & $0.789^{+0.011}_{-0.014}$ &$-0.18^{+0.167}_{-0.910}$ &$0.310^{+0.548}_{-0.185}$ \rule{0pt}{1.1em}  \\
\hline
$\Lambda$CDM& 2032.15 /1741 & 2036.15& 2047.076 & $0.2913^{+0.0012}_{-0.0012}$& $68.60^{+1.62}_{-1.59}$& - & - & -  \rule{0pt}{1.1em}  \\
\hline
  \hline \end{tabular}
\caption{Best fits, $\min\chi^2$, AIC and $\Delta$AIC from SNe Ia, $H(z)$, BAO and CMB
datasets for the EGB (\ref{Mod3}) and $\Lambda$CDM models.}
 \label{Estim}}
\end{table}
The EGB model (\ref{Mod3}) was analyzed in detail for all pairs of
$(\theta_j,\theta_k)$ its free model parameters (\ref{5param}),
these results are reproduced in Fig.~\ref{F3} in notation of
Fig.~\ref{F2}. One can note that the scenario is viable only with
small negative values for the parameter $\lambda$ in the range
$-4\cdot10^{-6}<\lambda<0$. This limitation and the condition
(\ref{dxi}) in the form $\dot\xi=\frac\lambda{\kappa^2H_0}a$ lead
to a weakly varying or nearly constant coupling factor $\xi$ with
the GB term. The term $8\lambda aE$ in equations (\ref{eq13}) and
(\ref{eq13}) appeared to be small ($|8\lambda aE|\ll 1$) for
$\lambda$ in the mentioned range and other parameters in the
limits from Table~\ref{Estim} at the late epoch $z<1000$ or
$a>10^{-3}$, because $E$ for viable solutions is close to the
$\Lambda$CDM expression (\ref{LCDM}) at this epoch:
$E\simeq\sqrt{\Omega_m^0}\cdot a^{-3/2}$. Hence, for $a>10^{-3}$
the value $aE$ has the upper bound of order
$\sqrt{\Omega_m^0}\cdot10^{3/2}$.

In the top-right panel in Fig.~\ref{F3} the evolution of the
scalar field $\varphi=\varphi(x)$ is depicted in the case of the
best fitted model parameters from Table~\ref{Estim}. We see that
the scalar field remains almost constant
$\varphi\simeq\varphi_0=0.31$ for redshifts $z<1000$ or $-6.9<x<0$
and deviates from its present value  $\varphi_0$ only at more
early times, when $\psi=H_0^{-1}\dot\varphi$ essentially deviates
from zero. Under these circumstances ($\varphi\simeq\varphi_0$,
$\psi\simeq0$, $|8\lambda aE|\ll 1$) at late times after the
recombination epoch the dynamical equation  (\ref{eq13}) becomes
close to  the $\Lambda$CDM equation (\ref{LCDM}), where the
parameter $\Omega_V$ plays the role of $\Omega_\Lambda$. However,
$\Omega_V$ (\ref{dimen}) in these EGB scenarios is a free model
parameter, the sum $\Omega_m^0+\Omega_V$ is not limited in
contrast to $\Lambda$CDM model. In particular, for the EGB model
(\ref{Mod3}) the sum $\Omega_m^0+\Omega_V$ of the best fitted
values from  Table~\ref{Estim} exceeds 1, it is true for $2\sigma$
CL domain in the $\Omega_m^0-\Omega_V$ panel in Fig.~\ref{F3}.
This freedom in choosing $\Omega_V$, obviously, leads to the
mentioned success of the model (\ref{Mod3}) in comparison with
$\Lambda$CDM model in minimizing $\chi^2$.
Additional degrees of freedom in the EGB scenario (\ref{Mod3}) lead to lower minimum of
$\chi^2$ and enlarges the best fit of $\Omega_m^0$ (with increased $1\sigma$ error
$\Delta\Omega_m^0$)  in comparison with $\Lambda$CDM model. The free model parameters
$\lambda$ and $\varphi_0$ are correlated it we search minimal values of $\chi^2$: if
$|\lambda|$ tends to zero, the corresponding fitted values of $\varphi_0$ also diminish.
The best fits of these parameters $|\lambda|\simeq1.8\cdot10^{-7}$  and
$\varphi_0\simeq0.31$ are attracted to zero borders of their domains in Fig.~\ref{F3}.

\section{Conclusions}
\label{Con}

In this paper, some scenarios of viable GW170817-compatible EGB
gravity have been studied and tested with observational data,
regarding their late-time phenomenology. These models include in
their Lagrangian a linear combination of the Ricci scalar and the
Gauss-Bonnet term particularly coupled to a scalar field $\phi$.
These models have been explored under the condition (\ref{ddxi})
on the coupling function $\xi(\phi)$ coming from coincidence of
the speed of gravitational waves  and the speed of light
\cite{Odintsov:2020sqy,Odintsov:2020zkl,Odintsov:2020xji,Oikonomou:2020oil}.
Scenarios of this type have been considered in the literature,
some of them can have suitable solutions  at late-times but with
singularities or instabilities at larger redshifts
\cite{OdintsovOF:2021,OdintsovSGS:2022EGB}.

In the first part of this paper, we have studied EGB models with
the gravitational action (\ref{action2}) where the Ricci scalar in
Lagrangian has a minimal coupling to the scalar field. We
considered scenarios with different types of the potential
$V(\phi)$, in particular, with power-law (\ref{Vpower}) and
exponential potential (\ref{Vexp}). These EGB scenarios have
solutions describing $H(z)$ and Pantheon+ SNe Ia observational
data (related to redshifts $0<z<2.5$) better than $\Lambda$CDM
model. However, these solutions for the potentials (\ref{Vpower}),
(\ref{Vexp}) and other viable potentials at high redshifts have
too quick growth of the Hubble parameter: $H(a)\sim a^{-3}$ (see
Fig.~\ref{F1}). Because of this asymptotic behavior we can not
describe satisfactory CMB and BAO observational data with the EGB
models coming from the action (\ref{action2}) under the condition
(\ref{ddxi}).

To avoid this drawback we have studied a more wide class of EGB
models determined by action (\ref{action}) and the constraint
(\ref{ddxi}). The model (\ref{Mod3}) from this class with the
factor $f(\phi)=\phi_0/\phi$ and the similarly behaving potential
$V(\phi)=V_0\phi_0/\phi$ appeared to be successful in tests with
all the observational data used in the paper: the Pantheon+ SNe Ia
data, $H(z)$ measurements, CMB and DESI BAO data. This model,
previously considered in papers
\cite{OdintsovOF:2021,OdintsovSGS:2022EGB}, includes solutions
with a regular at all redshift ranges considered. The model
(\ref{Mod3}) fits the mentioned observational data used
essentially better than $\Lambda$CDM model. The results of the
fittings comparison with $\Lambda$CDM model are summarized in
Table~\ref{Estim} and depicted in Figs.~\ref{F2}, \ref{F3}.

%%%%%%%%%%%%%%%%%%%%%%%%%%%%%%%%%%

To compare the EGB scenario (\ref{Mod3}) and $\Lambda$CDM model we
note that the absolute minima of the total $\chi^2$ function
(\ref{chitot}) radically differ: the EGB result
$\min\chi^2\simeq1997.76$ is lower than the $\Lambda$CDM value
$\min\chi^2\simeq2032.15$ with the large difference $34.39$, that
can be estimated as more than $3\sigma$ CL in Fig.~\ref{F2}. This
advantage of the EGB scenario (\ref{Mod3}) remains despite the
large number $N_p=5$ of its free parameters, when we consider
Akaike (\ref{AIC}) and Bayesian (\ref{BIC}) information criteria
\cite{Liddle_ABIC:2007}. The difference between results of these
models
$\Delta\mbox{AIC}=\mbox{AIC}_\mathrm{EGB}-\mbox{AIC}_\mathrm{{\Lambda}CDM}=-28.39$
and $\Delta\mbox{BIC}=-12.002$ remains large in favor of the EGB
model (\ref{Mod3}). Note that in our previous paper
\cite{OdintsovSGS:2022EGB} this EGB model did not have this
advantage: the AIC value for $\Lambda$CDM model was better than
for the EGB scenario. These changes are connected with the latest
SNe Ia Pantheon+ 1701 datapoints \cite{PantheonP:2022}, whereas in
Ref.~\cite{OdintsovSGS:2022EGB} we used 1048 datapoints of the
previous catalogue Pantheon \cite{Pantheon:2017}. This reason
leads to a similar advantage of the exponential $F(R)$ and some
other models if we compare them with  $\Lambda$CDM model
\cite{OdintsovSGS:2024}.

The best fits of free model parameters in Table~\ref{Estim} differ between two models
especially sharply (more than $3\sigma$) for the matter density parameter $\Omega_m^0$.
This difference for the Hubble constant $H_0$ is about $3\sigma$, but the EGB estimation
$H_0=65.90^{+1.48}_{-1.54}$ km\,/(s\,Mpc) is essentially lower.

Hence, we may conclude that the EGB model (\ref{Mod3}) under the important constraints
(\ref{ddxi}) on the speed of gravitational waves successfully fits well the latest
cosmological data, including Pantheon+ SNe Ia, $H(z)$, CMB and DESI BAO data. This EGB
model has an essential advantage when comparing with the $\Lambda$CDM model in
$\min\chi^2$ and if we use Akaike and Bayesian  information criteria.

\begin{acknowledgments}
This work was partially supported by the program Unidad de
Excelencia Maria de Maeztu CEX2020-001058-M, Spain (S.D.O).
\end{acknowledgments}

\end{document}